\documentstyle {article}

\begin{document}

{\bf Phase Analysis of Thin Film Oxide Systems by AES and EELS}

\

V.G. Beshenkov, V.A. Marchenko, V.T. Volkov and A.G. Znamenskii

\

Institute of Microelectronics Technology and High Purity Materials, Russian
Academy of Sciences, Chernogolovka, Moscow Region, 142432, Russia

\

$Abstract$

\

    Auger electron spectroscopy (AES) and reflection electron energy loss
spectroscopy (EELS in a reflection mode) are compared as the methods for
phase composition investigation of thin film oxide systems by ion
profiling. As an example, the $Al/Al_{2}O_{3}/Si$ and
$YBa_{2}Cu_{3}O_{7-x}/CeO_{2}/Al_{2}O_{3}$ systems are considered. The
adaptation of applied statistics methods for the AES and EELS data
treatment is discussed.

\

{\bf 1. Introduction}

\

    Both Auger and electron energy loss spectra display phase composition
of materials under study and can be obtained on standard electron
spectrometers. Lineshape and energy position sensitivity of Auger line to
chemical element bonding as well as electron energy loss structure near
elastic peak can be used as "finger prints" of uniphase materials. Due to
chemical reactions at the interfaces in thin film systems or surface
interaction with atmosphere, the mixtures of different phases occur in the
field of analysis and the spectra obtained are overlapped markedly. At the
same time, the phase composition information and the data treatment
principles are somewhat different in AES and EELS.

	 The present work is devoted to studying sharp interfaces (up to 5 nm),
when the mixing of spectra occurs through the ion beam etching ununiformity
and the existence of finite information length for escaping electrons.

\

{\bf 2. The method of phase analysis}

\

    Imagine a certain Auger spectrum or electron energy loss spectrum in a
multidimensional linear vector space. Take the number of coordinate axes
equal to the number of accumulated intensities {\bf m} in the spectrum and
specify coordinates by corresponding intensities. The point obtained is an
image of isolated spectrum and represents the object of  classification
(ordination), when the set of {\bf n} spectra is mapped. The sequence of
points is consistent with the depth profile and reflects the alteration of
phase composition in depth. Due to a partial duplication of the information
contained in spectra the geometric configuration of image points can be
displayed in a space of low dimensionality. In order to transit into such a
space with the lowest distortion of the angles and distances between the
points, the methods of reduction of space dimensionality as principal
component analysis or factor analysis can be used [1-3]. In particular, the
analyzed data can be projected into the spaces defined by one, two, three,
etc. principal components. There exist some tests of applied statistics to
define the dimensionality of such a space for which the loss of information
by data projection occurs at the cost of random experimental error [2-4].
The most convenient test, to our mind, is the graphical method based on an
equivalence of secondary ("noise") eigenvectors in an eigendecomposition of
covariance matrix [4]. We decompose ${\bf n} \times {\bf n}$ covariance
matrix ${\bf S=X^{T}X}$ ({\bf X} - ${\bf m} \times {\bf n}$ data matrix,
${\bf m>n}$) and use {\bf n} column elements ${\bf h_{ij}}$ {\bf i=1,...n
j=1,2} of the two first eigenvectors as spectra coordinates in the
two-dimensional case.  The elements ${\bf h_{ij}}$ {\bf i=1,...n j=1,3} are
additionally used to construct the orthogonal plane in the
three-dimensional case. It is quite natural to connect the dimensionality
of the space with the phase number in the object under study.

The inherent AES ability to sense Auger lines of different chemical
elements enables one to find "uniphase" regions in depth distribution by
principal component analysis and to use them for phase concentration
calculations, including mixing with the phases unanalyzable on the
Auger line under study. One needs to select in the projection map such
types of point configurations as point clusters or points on a straight
line which contains the origin of coordinates. The objects of
classification (spectra in the set) corresponding to that points may be
considered as phase references, because the loading coefficients on the
phase are unity or can be scaled to unity. Other phases, at the same
time, have zero loadings. The loading coefficients deduced by oblique
rotation of principal axes for the points composing the interface can
be defined as phase concentrations (fractions) at the interface, if the
axis scaling is implemented in appropriate manner (new axes are
directed to phase references).

In the case of the absence of uniphase
	 regions in depth profile defined by Auger line projecting, there is a
	 need to use the references of chemical compounds. When all the phases
	 are identified in accordance with the number of the phases deduced
	 earlier, the phase concentrations can be obtained by the multiple
	 linear regression method or target transformation procedure [3-4]. The
	 statistical method of phase identification using the references based
	 on the hypothesis test is presented in [5].

 One can obtain the
	 statistically unsatisfactory agreement of the oxide references with the
	 oxide system under study because a high energy and high current
	 electron beam (up to $1 \mu A$ and 10 keV) are used in AES to excite
Auger spectra which can produce radiation damage of oxide components. EELS
	 permits one to decrease substantially the electron beam current and
	 energy (up to 10 nA, 500 eV). Unfortunately, the element
	 characteristicity in EELS is broken, so the references of all the
	 phases are required simultaneously for concentration calculations. Some
	 difficulties in the EELS phase analysis can be caused by presumed
	 alteration of surface plasmon energies with the alteration of
	 dielectric constant of the media during depth profiling of an
	 interface, for example, a dielectric-metal interface. The decrease
	 of plasmon excitation probability in ultrathin layers or near
interfaces can initiate some efforts to scale the phase concentrations
obtained.

We employ the biplot technique using row coordinates of the {\bf H}
	 and {\bf XH} matrices balanced by appropriate singular value
	 multiplication to project EELS spectral intensities of the whole
	 spectra set into the same space, where the EELS spectra are mapped [1].
	 If we connect the {\bf n} spectra points of depth profile with the
	 origin of coordinates, it can be seen that some parts of the {\bf m}
	 spectral intensity points lie near the vectors directed to clusters,
	 trends, etc. of the spectra set. It means that such spectral
	 intensities have significant loadings onto the corresponding spectra.
	 The aim of this study was to define the spectral intensity intervals
	 responsible for the occurrence of specific interfacial components in
	 the case of sharp interphase boundaries.

\

{\bf 3. Experimental}

\

    The depth profiling was carried out by successive cycles of ion
	 sputtering and data accumulation on the Auger microprobe model
	 JAMP-10S. A primary electron beam of 10 keV and 1 $\mu A$ (defocused up
	 to 100 $\mu m$ to minimize radiation damage) was used for Auger and of
	 400 eV and 10 nA for EELS spectra excitations, respectively. The Auger
	 spectra were measured in the derivative mode $E \times N'(E)$ with a
	 modulation amplitude of 5 V. Accumulation time was 250 ms/eV with the
	 energy step of 1 eV. To avoid the charging of $Al_{2}O_{3}$ wafers, the
	 normal beam incidence was changed to $45^{0}$ incidence by rotating the
	 sample plane and the beam energy was lowered to 5 keV. The EELS spectra
	 were measured in the integral mode $E \times N(E)$. The accumulation
time was 500 ms/eV with the energy step of 0.5 eV. The energy resolution
	 was about 0.01. 1 keV or 3 keV $Ar^{+}$  ion bombardment was used in
	 accordance with the film thickness. As a rule, the sets of 50 spectra
	 with 80 intensity points in a spectrum were representative. To exclude
	 constant background components, the spectra were centered [1].

	 The Auger and EELS spectra sets of the $Al/Al_{2}O_{3}/Si$ and
	 $YBa_{2}Cu_{3}O_{7-x} /$ $CeO_{2} /Al_{2}O_{3}$ thin film oxide systems
	 obtained in the coarse of ion sputtering were investigated using
	 applied statistics methods. The characteristic features of the systems
	 under study are both simple structure of losses and well-defined energy
	 difference between principal loss peaks of contacting materials. To
	 minimize the interaction of film layers coming in contact, the
	 appropriate regimes of film production and heat treatment were used.
	 The measured energy positions of peaks were in error by 0.5 eV.

\

{\bf 4. Results and discussion}

\

    1.1. The $Al/Al_{2}O_{3}$  interface. Some of Auger spectra of  the
	 $Al/Al_{2}O_{3}$ system (the O KLL Auger line) are presented in Fig.1. The
	 dimensionality of the data space was two. With the rise of sputter time
	 the projections of spectra into the space of the two first principal
	 components are arranged (Fig.2) in the origin of coordinates (zero
	 level of O KLL intensity), then on the line directed to the origin of
	 coordinates (the uniphase region of some oxygen-containing phase in the
	 presence of unanalyzable phase of $Al$) and after that they turn out to
	 the point cluster corresponding to $Al_{2}O_{3}$ film. The composition
	 profiles of oxygen-containing phases obtained by oblique rotation of
	 coordinates are presented in Fig.3. The first phase corresponds to the
	 initial state of $Al_{2}O_{3}$, the other corresponds to the
	 same $Al_{2}O_{3}$ phase destroyed by electron and ion beams. It
	 should be noted that the mixture of phases at the interface in this
	 case is not the mixture of real phases of a constant composition due to
	 the ability of AES to sense a nearneighbourhood of exited atom.

The EELS spectra of the $Al$ film, the $Al/Al_{2}O_{3}$ interface and the
	 $Al_{2}O_{3}$ film are presented in Fig.4.  The energies of bulk $Al$
	 plasmons (the first and the second bulk plasmons are seen at the
	 energies of $\Delta E = 14.5 eV$ and $\Delta E = 30 eV$, the surface
	 plasmons are not significant under these experimental conditions and
 are seen as high energy peak shoulders) correspond to the literature data.
The EELS spectra of $Al_{2}O_{3}$ contain one wide peak corresponding to
	 interband transitions. The dimensionality of the data space set by the
	 EELS spectra of depth profile including the interface was three. It
	 presumes the existence of an appreciable specifical interfacial
	 component. The biplot technique shows in the plane of the first and the
	 second principal axes (Fig.5) that the tendency of point displacement
	 along the line with sputter time increasing is attributed to a
	 permanent change of contributions of spectral features with the
	 energies of $\Delta E = 14.5 eV$ (and also 30 $eV$) and $\Delta E = 21
eV$ in the transition region between $Al$ and $Al_{2}O_{3}$. This fact does
not contain any new information. However, it is seen in the plane of the
first and the third principal axes (Fig.6) that the rise of interfacial
	 component content (maximum contents is observed at the sputter time of
	 11 min.) is attributed to spectral features with the energies of
	 $\Delta E = 6.5 eV$ and $\Delta E = 18 eV$ occurred simultaneously at
	 the interface. These features are not seen in Fig.4, but can be seen in
	 the difference spectrum (Fig.7), that is the difference between the 11
	 min.  spectrum and the linear combination of the $Al$ and $Al_{2}O_{3}$
	 spectra composed with appropriate weights.  The contribution of these
	 features into the interface spectrum is about 0.05 of spectrum
	 intensity. The energies of the features obtained by the biplot
	 technique are exactly equal to the difference and half-sum of the
	 energies of principal spectral features of $Al$ (bulk plasmon) and
	 $Al_{2}O_{3}$ (interband transitions) films coming in contact.  As can
	 be seen further this phenomenon is registered on a great variety
	 of interfaces which corresponds to both simple structure of losses and
	 well-defined energy difference of principal loss peaks of contacting
	 materials. We can furnish an explanation of it as the interaction of
	 collective excitations (and interband transitions in a particular case)
	 at the analyzed interface under electron beam, that is the adventition
	 of the beats in a system of harmonic oscillators (for example,
oscillations which take place with some frequency corresponding to the
energy of $\Delta E = 14.5 eV$) under an external periodic force ($\Delta E
= 21 eV$). Earlier the occurrence of a peak with the energy of $\Delta E =
7.1 eV$ in the EELS spectra of ultrathin $Al_{2}O_{3}$ films obtained by in
situ oxidation of $Al$ was observed [6].  Phenomenological dielectric
theory was used in [7] to interpret this peak as a modified $Al$ surface
plasmon peak shifted on an energy scale due to the alteration of dielectric
constant of the media.  The proposal about the existence of a surface
plasmon excitation with the energy of $\Delta E_{s}=((\Delta E^{2}+\Delta
E'^{2})/2)^{1/2}$, where $\Delta E$ and $\Delta E'$ are the bulk plasmon
energies of contacting materials (metals), was also suggested in [7].
Despite of the fact that this value is close to the value of a half-sum of
the energies of the principal loss peaks obtained, we believe that there is
another physical phenomenon. Both the constancy of "modified" plasmon
energy at the interface region with varied phase composition and the
simultaneous occurrence of two types of energy losses corroborate our
conclusion.

1.2.  To minimize the electron and ion beam radiation effects
	 on the interface EELS spectra registration, the model system $Sn/Si$ was
	 studied. The $Sn$ film was obtained at liquid nitrogen temperature by rf
	 magnetron deposition on the $Si$ wafer cleaned in rf plasma to remove a
	 native oxide. There was no $Sn$ continuous coating, but we believe that
	 the interface is sharp and the diffusion interaction of $Sn$ and $Si$ is
	 negligible. The EELS spectra of the $Sn$ film and the $Si$ wafer with
their interface  obtained in the course of depth profiling are presented in
	 Fig.8. The system under study was more complex than the previous one
	 due to a high dimensionality of the data space (attributed to the
	 surface plasmon intensity alteration) but it is seen (Fig.9) that the
	 specific spectral feature with the energy of $\Delta E = 14.5
eV$ equal to the half-sum of the energies of the
principal loss peaks ($\Delta E = 12.5 eV$ and $\Delta
	 E = 16.5 eV$) of $Sn$ and $Si$ also exists.  The
	 feature with a difference energy comes into elastic peak region.

1.3. The $Bi/Si$ system was investigated to confirm the results obtained
	 earlier. The absence of any new phase at film-substrate interface was
	 confirmed by Auger depth profiling. The EELS spectra of the $Bi$ film,
	 the $Si$ substrate and their interface are presented in Fig.10. The
	 specific interface features are seen in Fig.11: with the energy of
	 $\Delta E = 7 eV$ corresponding to the difference of $\Delta E = 16.5
	 eV(Si)$ and $\Delta E = 9.5 eV(Bi)$, with the
	 energy of $\Delta E = 19 eV$ corresponding to the
	 difference of $\Delta E = 33 eV(Si)$ and $\Delta
	 E = 14 eV (Bi)$, with the energy of $\Delta E =
	 23 eV$ corresponding to the difference of $\Delta E
	 = 33 eV(Si)$ and $\Delta E = 9.5 eV(Bi)$
	  and, also, to half-sum of $\Delta E = 33 eV(Si)$
	  and $\Delta E = 14 eV(Bi)$.  The
	 dimensionality of the data space was three that points to the existence
	 of unique interfacial component with 7, 19 and 23 eV features.

	 2. The $YBa_{2}Cu_{3}O_{7-x}/CeO_{2}/Al_{2}O_{3}$ system.

	 2.1. The
	 $YBa_{2}Cu_{3}O_{7-x}/CeO_{2}$ interface. Some content of a new phase
	 (up to 0.3, connected with thermal decomposition of
	 $YBa_{2}Cu_{3}O_{7-x}$) was detected at the interface by Auger depth
	 profiling (Fig.12). The identity of phase composition profiles obtained
	 by the Cu LMM and Ba MNN Auger line studies supports the assumption
	 that the existence of a new phase mixture is also possible. Electron
	 and ion beams in the AES mode modifies the $CeO_{2}$ film not only at
	 the interface but also in the $CeO_{2}$ film bulk as can be seen from
	 the Ce MNN Auger line study (the $CeO_{2}$ concentrations were plotted
	 on the base of p-to-p Ce MNN intensities for the sputter times higher
	 than 15 min.).  It should be mentioned that the existence of a real new
	 phase at the $YBa_{2}Cu_{3}O_{7-x}/CeO_{2}$ interface is not an authentic fact due to
	 the absence of the data for higher temperature depositions of the
	 $YBa_{2}Cu_{3}O_{7-x}$ film and small thickness of the interfacial layer as
	 compared with the depth resolution. There is a
	 mismatch of phase concentrations of $CeO_{2}$ obtained by the analysis
	 of the Ce MNN Auger inelastic losses in the Ba MNN Auger line region and
	 the $CeO_{2}$ concentrations obtained by the analysis of  the Ce MNN
	 Auger line. Therefore, the study of electron energy losses is important
	 in AES.

	 Despite of possible existence of a new phase at the interface, the EELS
	 spectra study was performed. The EELS spectra of the
	 $YBa_{2}Cu_{3}O_{7-x}$ film (principal features are with the energies
	 of $\Delta E = 12.5 eV$ and $\Delta E = 23 eV$
	 ), the $CeO_{2}$ film (principal features are with
	 the energies of $\Delta E = 13 eV$ and $\Delta E =
	 28.5 eV$) and their interface are presented in
	 Fig.13. The specific interfacial features of low intensities with the
	 energies equal to the difference and half-sum of the energies of
	 principal loss peaks with the energies of  $\Delta E = 12.5 eV$
	 and $\Delta E = 28.5 eV$ are designated
	 in Fig.14 by arrows.

	 2.2. The $CeO_{2}/Al_{2}O_{3}$ interface.  There are (Fig.15-16)
	 specific interfacial features of low intensities with the energies of
	 $\Delta E = 8 eV$ and 16.5 eV equal to the difference and half-sum of
	 the energies of principal loss peaks with the energies of $\Delta E =
	 13 eV (CeO_{2})$ and $\Delta E = 21 eV (Al_{2}O_{3})$.  The
	 concentration profiles of $Al$-containing phases were identical with
	 those presented in Fig.3.

\

{\bf 5. Conclusions}

\

    The AES depth profiling study of thin oxide film systems indicates that
	 there always exists an interfacial oxide layer where the modification
	 of composition by electron and ion beams occurs. The EELS spectra are
	 not sensitive to this process. On the other hand, the specific
	 interfacial features corresponding to a distinct interaction of
	 collective excitations (and interband transitions) at the interphase
	 boundaries of contacting materials occur in the absence of a new phase
	 formation. This fact appreciably lowers the analytic usefulness of EELS
	 as a method of phase analysis. The interaction defined is responsible
	 only for a few percents of spectral intensities of the analyzed
	 interface in its maximum and is not of long-range order.

	 The work was
	 supported by Russian Foundation for Basic Research, grant 97-03-33557.

\

{\bf References}

\

1.  Handbook of Applicable Mathematics. Volume VI: Statistics.
	  Part B.  / Eds. W. Ledermann and E. Lloyd. - N.Y.: Wiley, 1984.

2.  $Aivazyan \ S. A., \ Buchstaber V. M., \ Yenyukov \ I. S., Meshalkin \
	 L. D.$  Applied Statistics:  classification and reduction of
	  dimensionality. - Moscow: Finansy i statistika, 1989.

3.  $Malinowski \ E. R., \ Howery \ D. J.$ Factor Analysis in Chemistry. -
	  N.Y.:  Wiley, 1980.

4.   $Beshenkov \ V. G., \ Kopetskii \ Ch. V., \ Shiyanov \ Yu. A.$ Phys.
	  Stat.  Sol. (a).  1989. V. 114.  P. 191.

5. $Beshenkov \ V. G., \ Znamenskii \ A. G., \ Marchenko \ V. A.$ Izvestiya
	  RAN. Ser.  fiz. 1998.  V. 62. No 3. P. 517.

6. $Powell \ C. J., \ Swan \ J. B.$ Phys. Rev. 1960. V. 118. P. 640.

7. $Stern \ E. A., \ Ferell \ R. A.$ Phys. Rev. 1960. V. 120. P. 130.

\

FIGURE CAPTIONS

\

Fig. 1. The derivative O KLL Auger spectra of $Al/Al_{2}O_{3}$ . The values
	  of sputter time (in min.) corresponding to the spectra are designated.

Fig. 2. The projections of the O KLL Auger spectra of $Al/Al_{2}O_{3}$  onto
	  the plane of the first and the second principal components. The values
	  of sputter time (in min.) corresponding to the spectra are designated.

Fig. 3. The composition profiles of oxygen-containing phases of
	  $Al/Al_{2}O_{3}$ in depth:  the initial $Al_{2}O_{3}$ phase (circles)
	  and the $Al_{2}O_{3}$ phase modified by electron and ion beams (squares).

Fig. 4. The EELS spectra of $Al/Al_{2}O_{3}$: the $Al$ film (circles), the
	  $Al_{2}O_{3}$ film (squares) and the $Al/Al_{2}O_{3}$  interface
	  (triangles).

Fig. 5. The $Al/Al_{2}O_{3}$ biplot map: the EELS spectra in depth (squares), the
	  EELS spectral intensities of the whole depth profile (circles)
	  projected onto the plane of the first and the second principal axes
	  (j=1,2). The values of sputter time (in min.) corresponding to the
	  spectra and the energies of principal features are designated.

Fig. 6. The $Al/Al_{2}O_{3}$ biplot map : the EELS spectra in depth (squares), the
	  EELS spectral intensities of the whole depth profile (circles)
	  projected onto the plane of the first and  the third principal axes
	  (j=1,3). The values of sputter time (in min.) corresponding to the
	  spectra and the energies of principal features are designated.

Fig. 7. The EELS spectrum of $Al/Al_{2}O_{3}$ (up triangles) and its projection
	  onto the plane of the first and the second principal components (down
	  triangles) and the difference spectrum (stars).

Fig. 8. The EELS spectra of $Sn/Si$:  the $Sn$ film (circles), the $Si$
	  substrate (squares) and the $Sn/Si$ interface (triangles).

Fig. 9. The $Sn/Si$ biplot map:  the EELS spectra in depth (squares),  the
	  EELS spectral intensities of the whole depth profile (circles)
	  projected onto the plane of the first and the third principal axes
	  (j=1,3). The values of sputter time (in min.) corresponding to the
	  spectra and the energies of principal features are designated.

Fig. 10. The EELS spectra of $Bi/Si$: the $Bi$ film (circles),  the $Si$
	  substrate (squares) and the $Bi/Si$ interface (triangles).

Fig. 11. The $Bi/Si$ biplot map: the EELS spectra in depth (squares), the
	  EELS spectral intensities of the whole depth profile (circles)
	  projected onto the plane of the first and the third principal axes
	  (j=1,3). The values of sputter time (in min.) corresponding to the
	  spectra and the energies of principal features are designated.

Fig. 12. The composition profiles of $YBa_{2}Cu_{3}O_{7-x}/CeO_{2}$ in
	  depth:  the initial $YBa_{2}Cu_{3}O_{7-x}$ phase (circles), the
	  initial $CeO_{2}$ phase (down triangles), the $CeO_{2}$ phase modified
	  by electron and ion beams (squares) and the new phase at the interface
	  (up triangles).

Fig. 13. The EELS spectra of $YBa_{2}Cu_{3}O_{7-x}/CeO_{2}$: the
	  $YBa_{2}Cu_{3}O_{7-x}$ film (circles), the $CeO_{2}$ film (squares)
	  and the $YBa_{2}Cu_{3}O_{7-x}/CeO_{2}$ interface (triangles).

Fig. 14. The $YBa_{2}Cu_{3}O_{7-x}/CeO_{2}$ biplot map: the EELS spectra
	  in depth (squares), the EELS spectral intensities of the whole depth
	  profile (circles) projected onto the plane of the first and the third
	  principal axes (j=1,3). The values of sputter time (in min.)
	  corresponding to the spectra and the energies of principal features
	  are designated.

Fig. 15. The EELS spectra of $CeO_{2}/Al_{2}O_{3}$: the $CeO_{2}$ film
	  (circles), the $Al_{2}O_{3}$ substrate (squares) and the
	  $CeO_{2}/Al_{2}O_{3}$ interface (triangles).

Fig. 16. The $CeO_{2}/Al_{2}O_{3}$ biplot map : the EELS spectra in depth
	  (squares), the EELS spectral intensities of the whole depth profile
	  (circles) projected onto the plane of the first and the third
	  principal axes (j=1,3). The values of sputter time (in min.)
	  corresponding to the spectra and the energies of principal features
	  are designated.

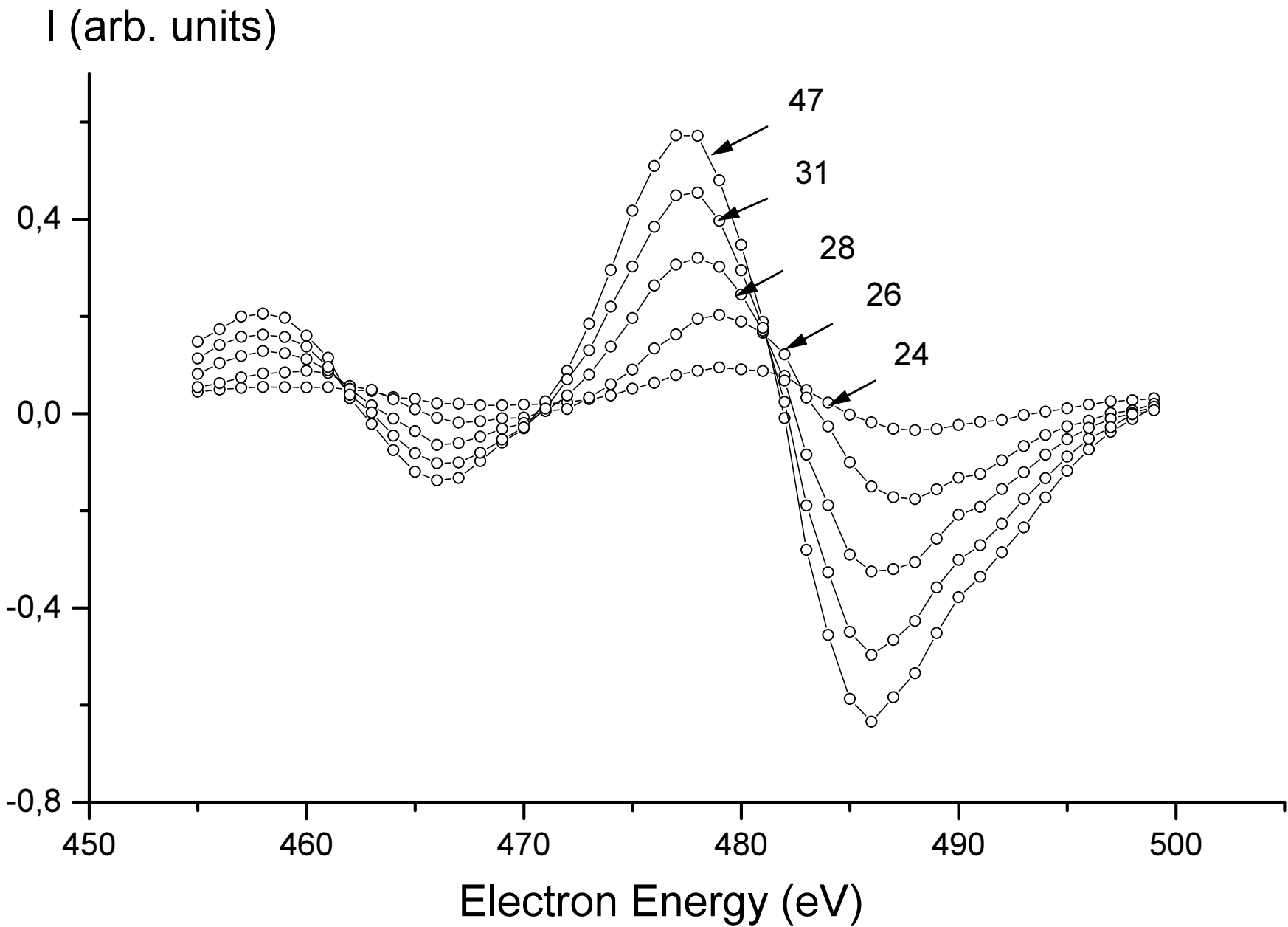
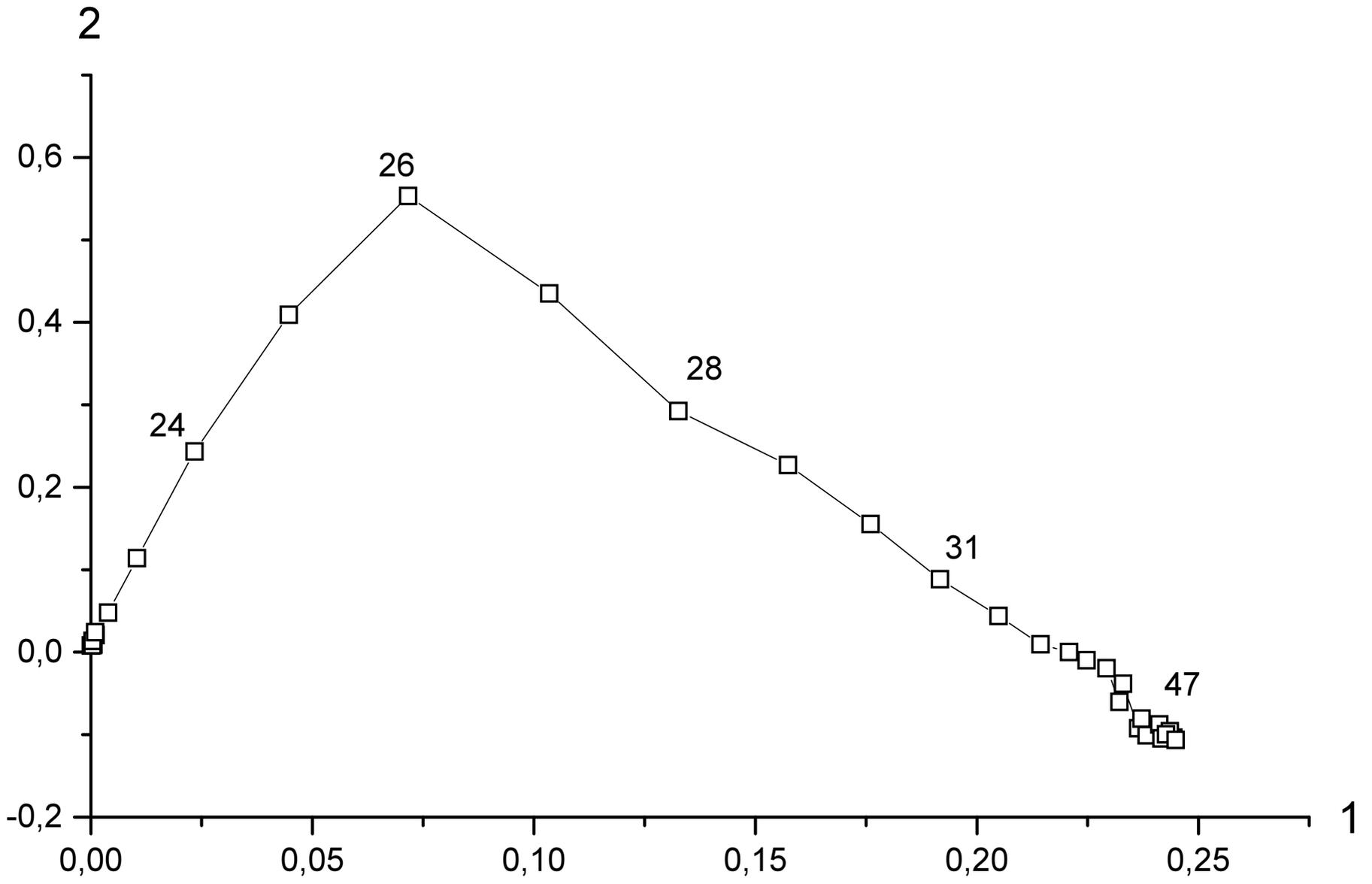
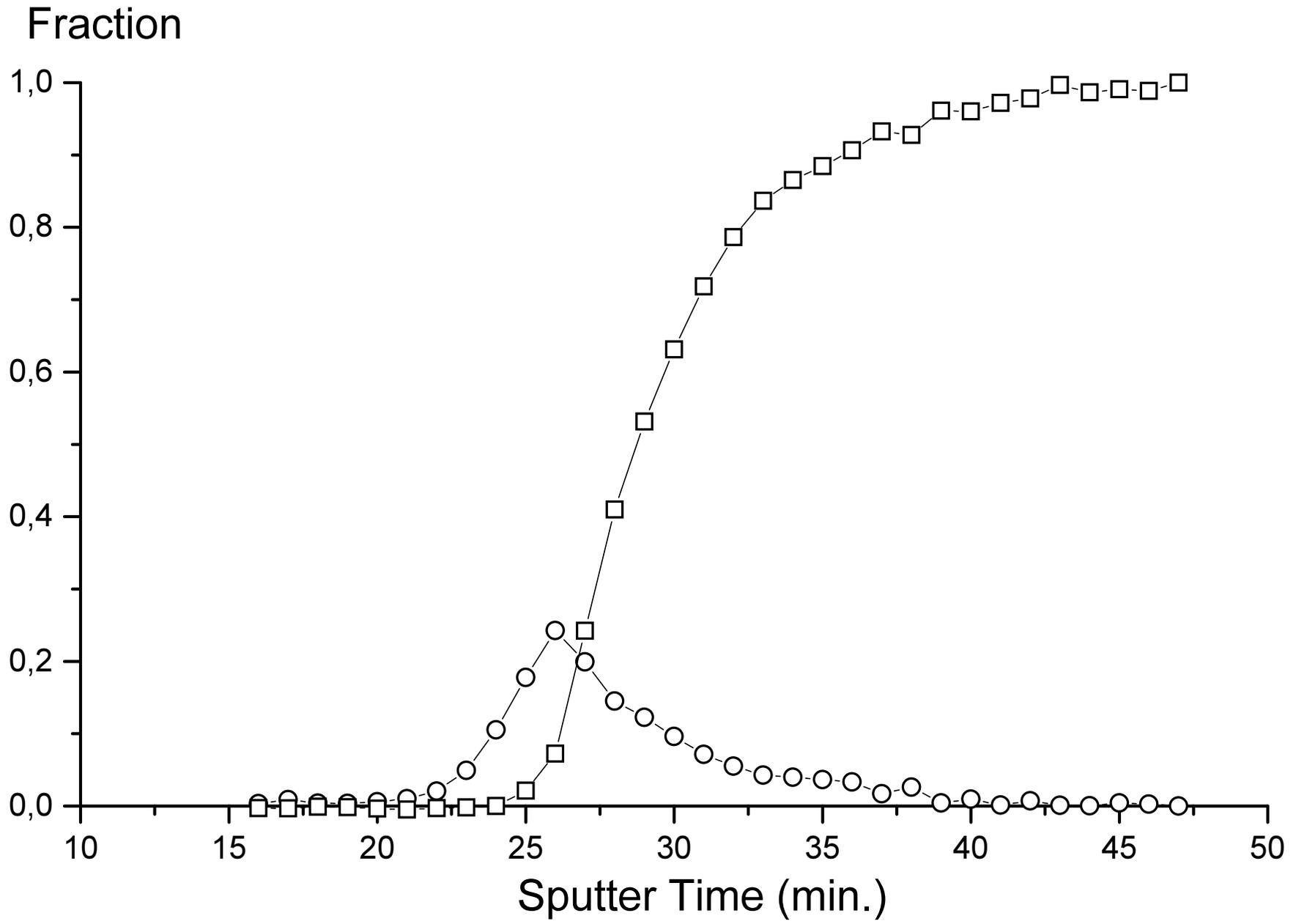
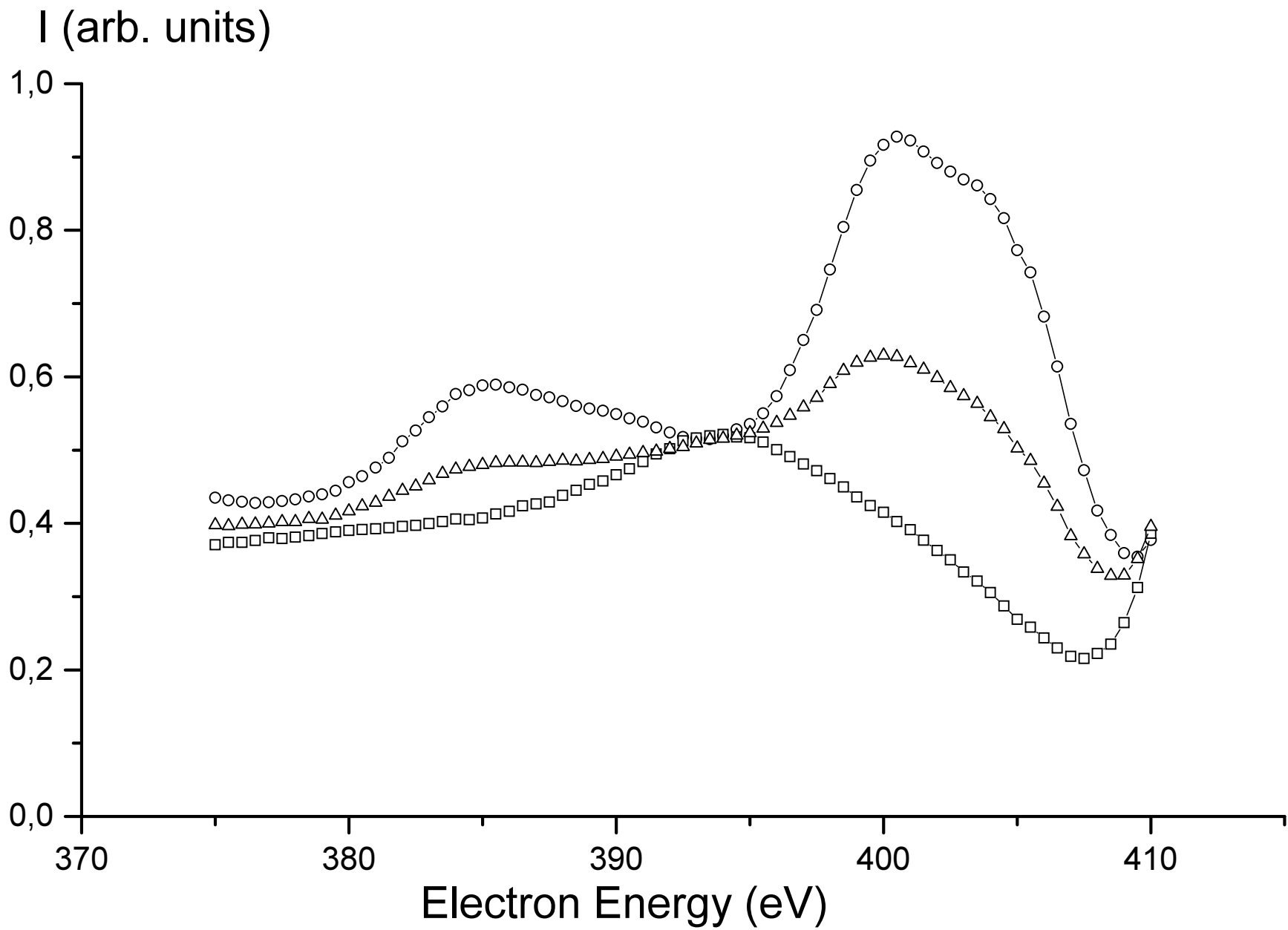
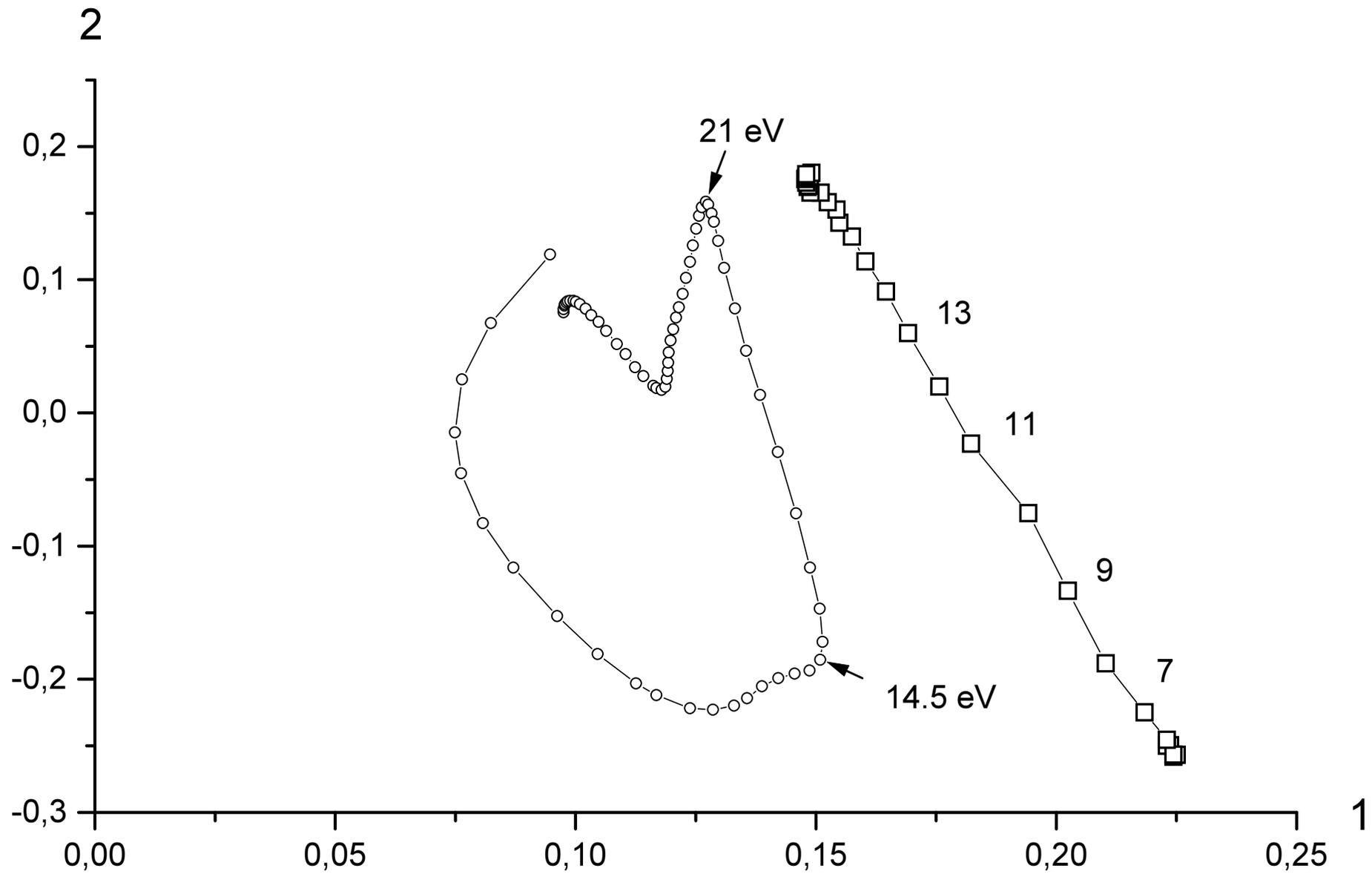
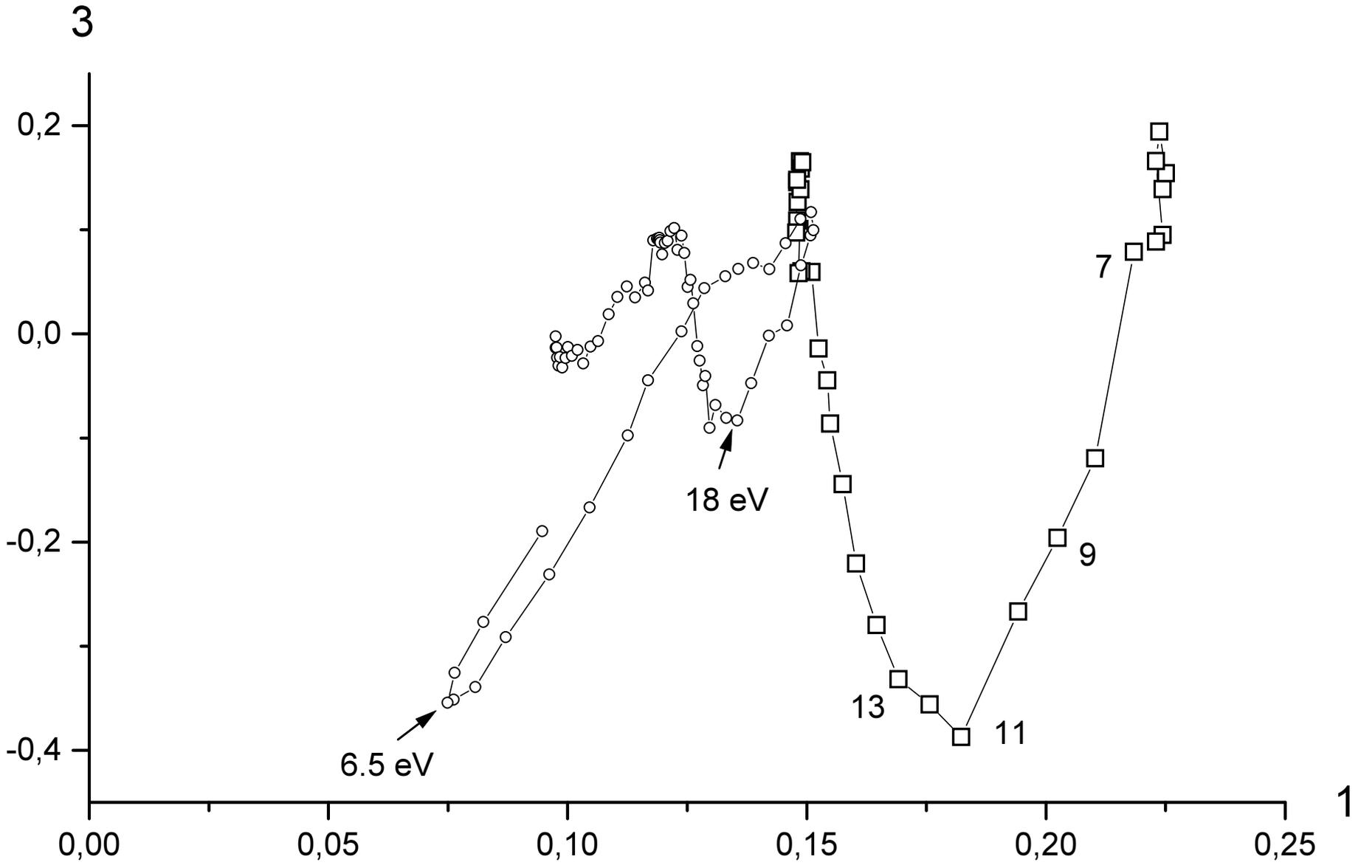
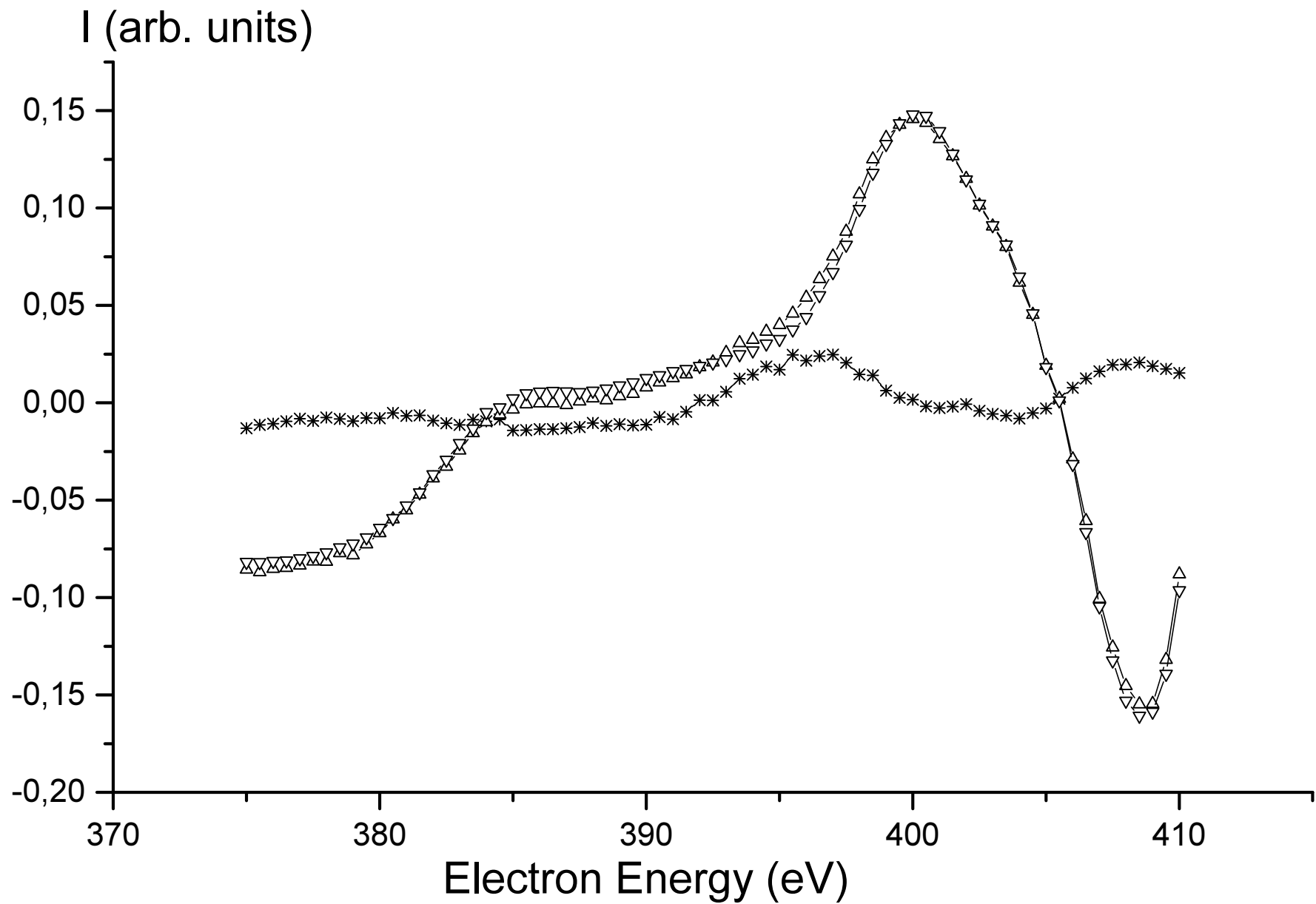
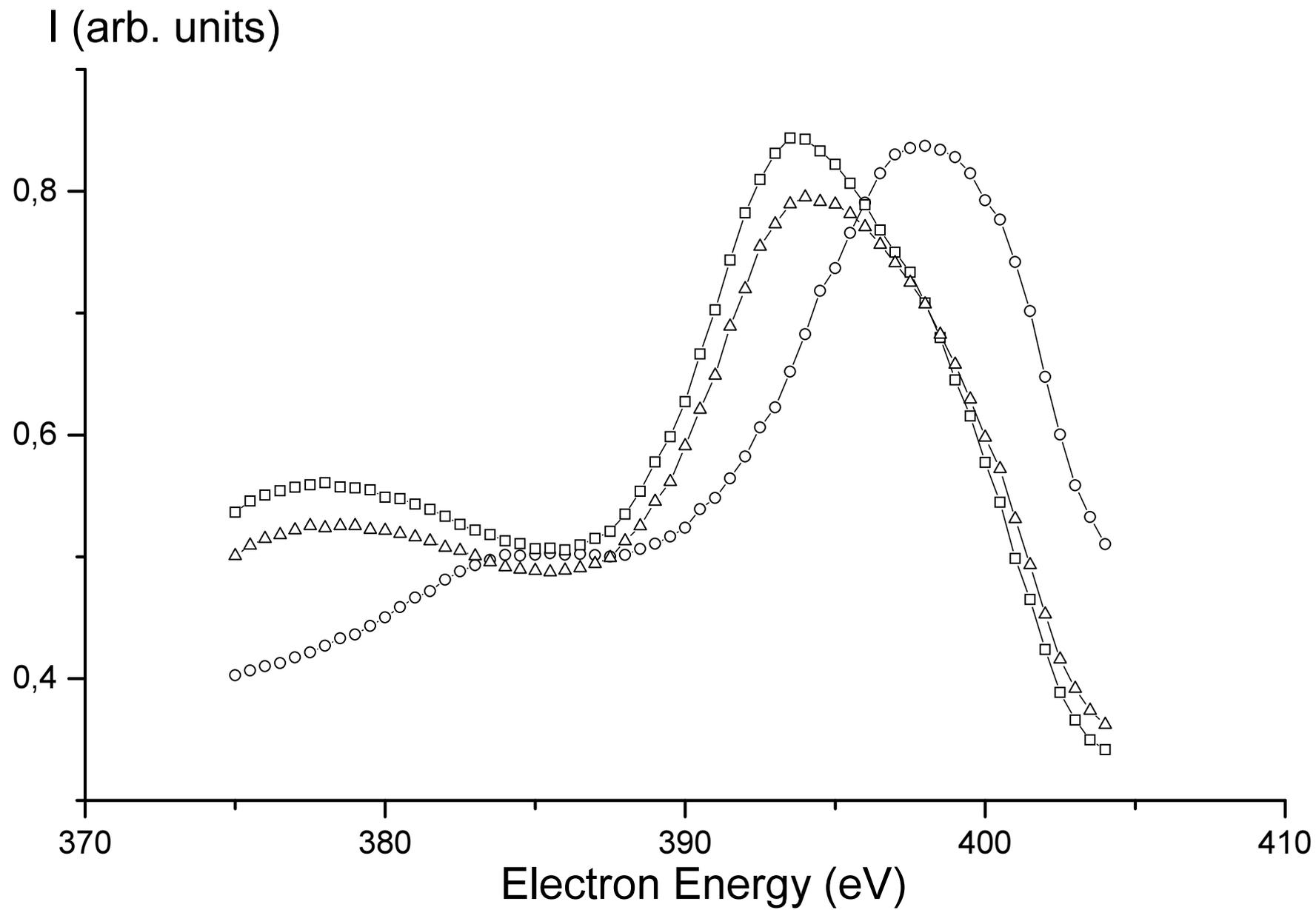
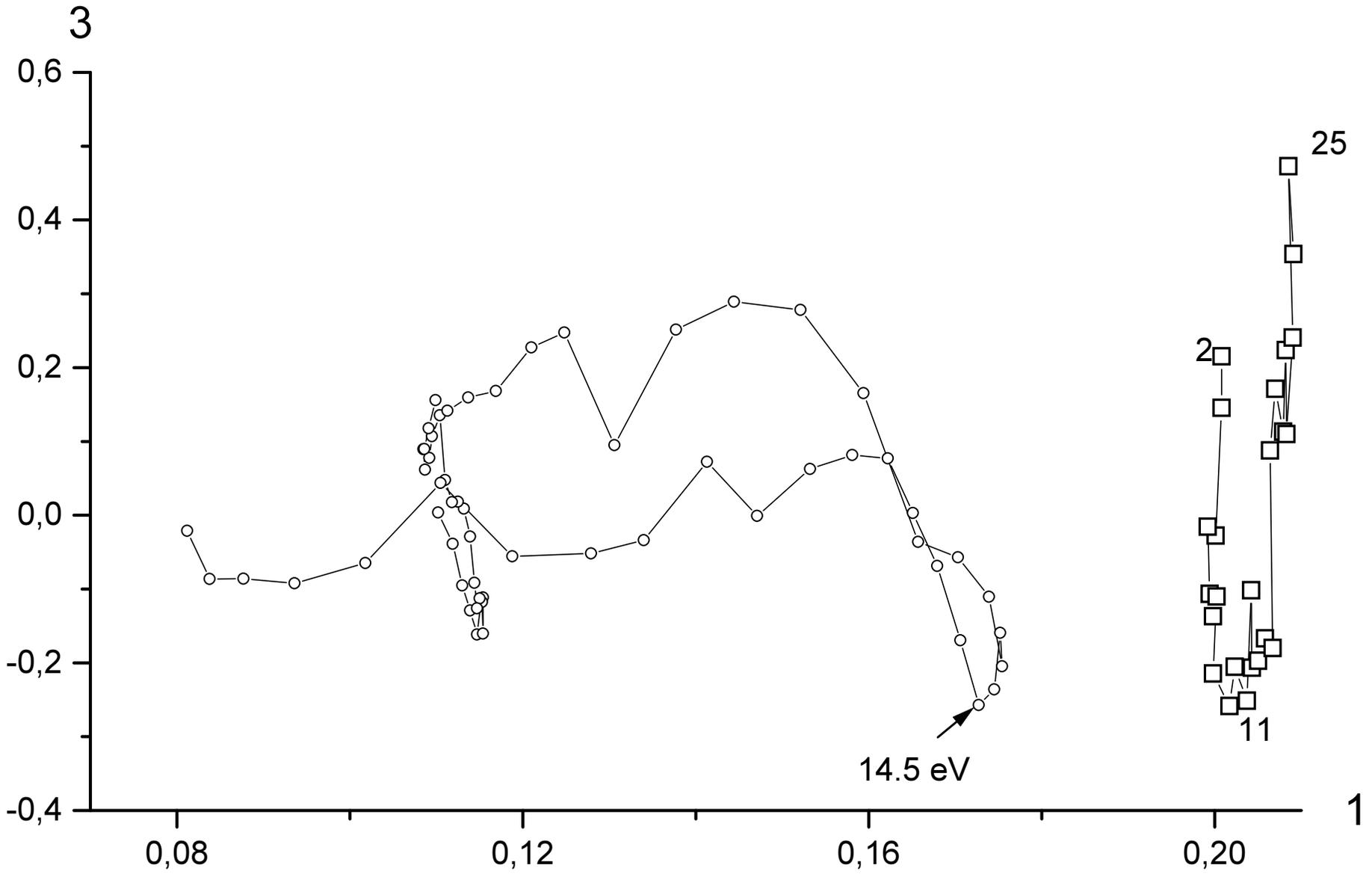
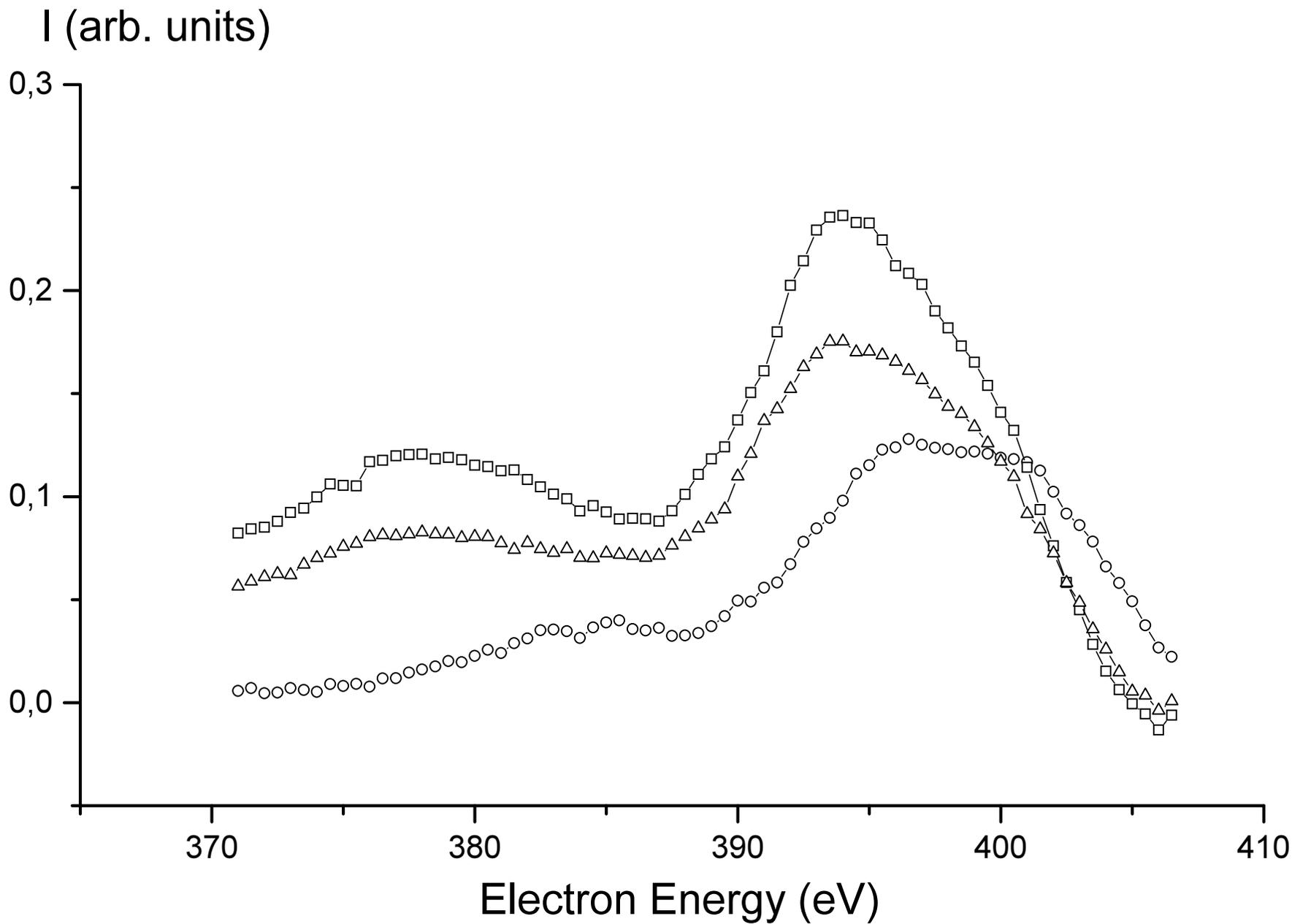
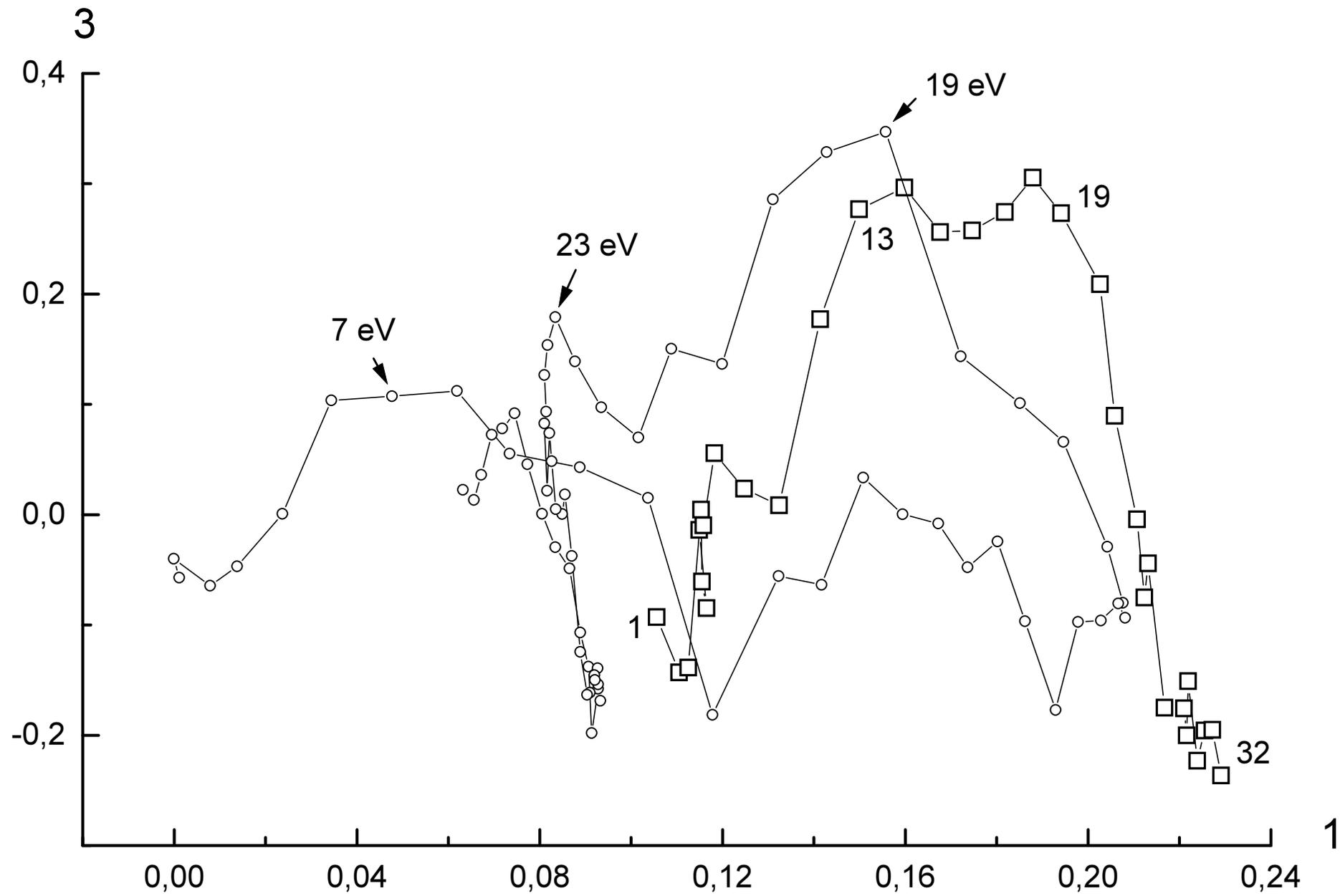
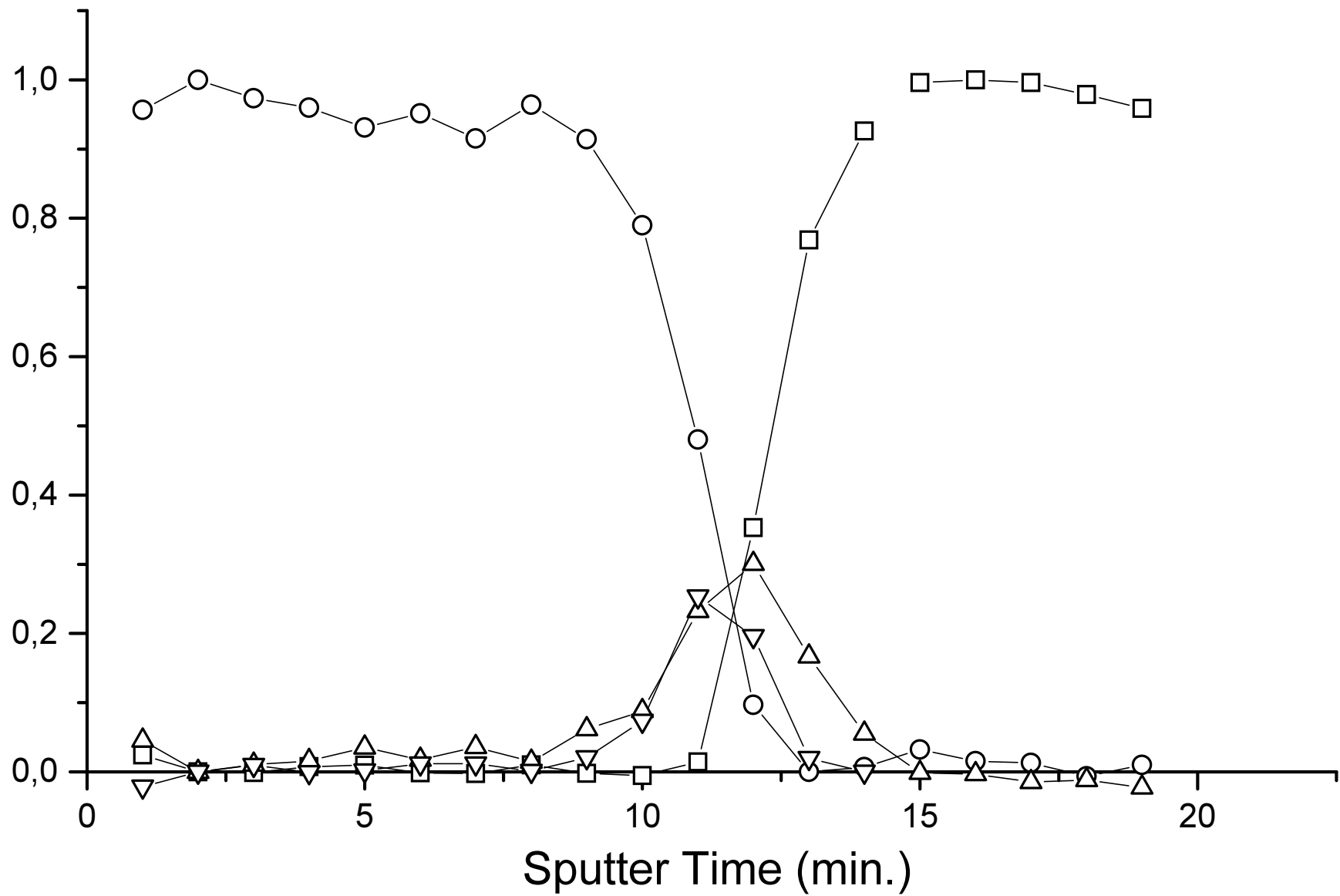
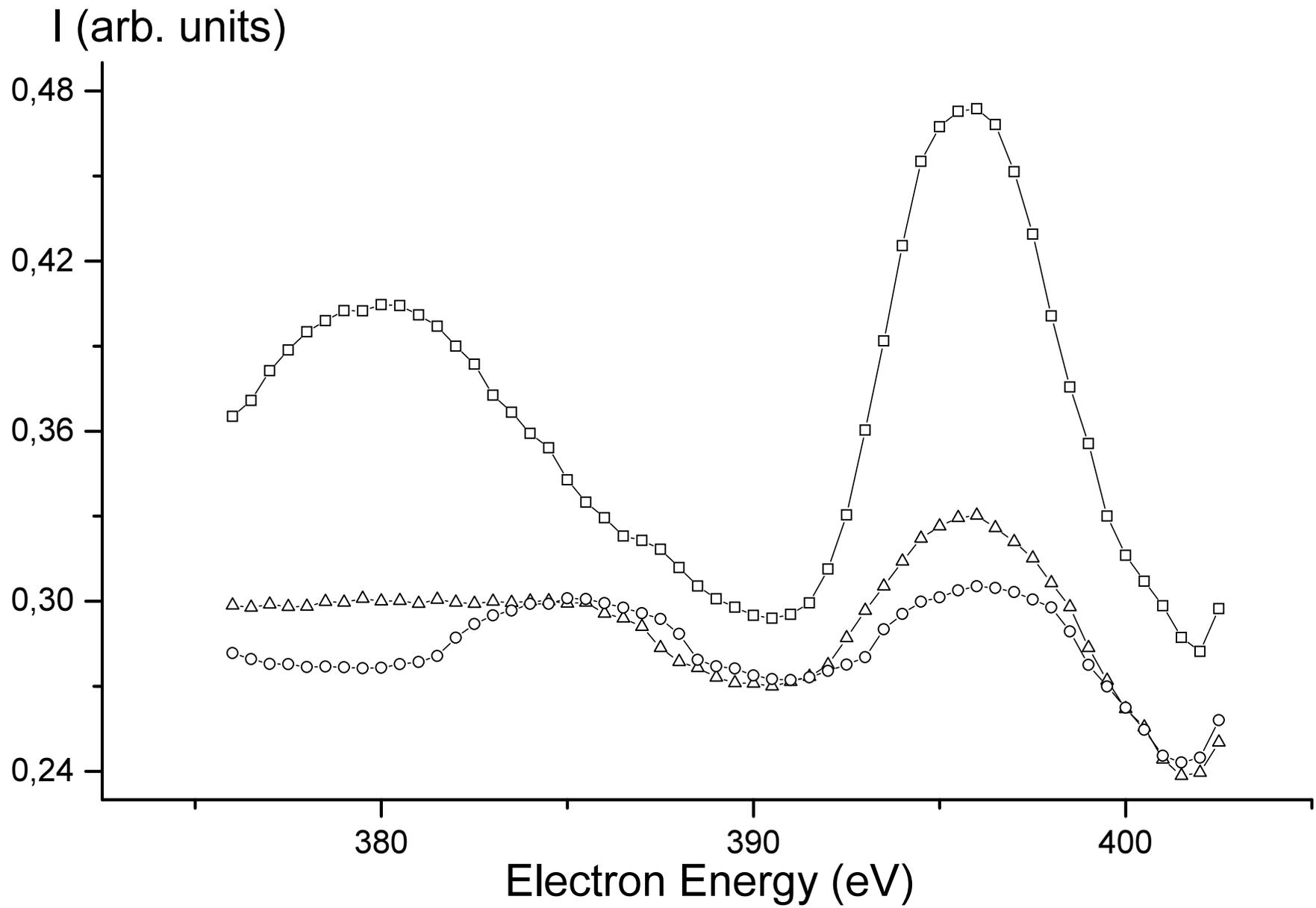
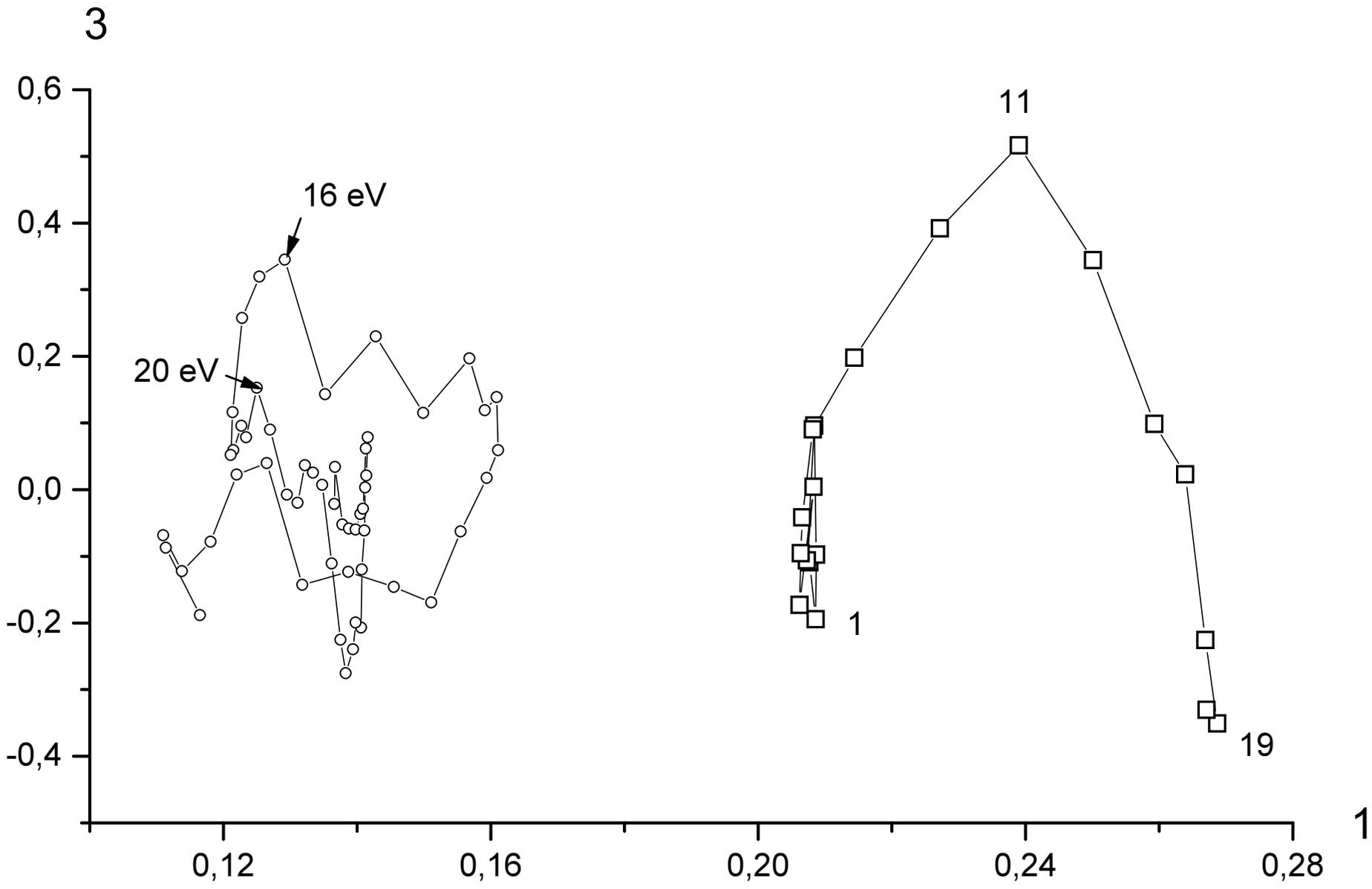
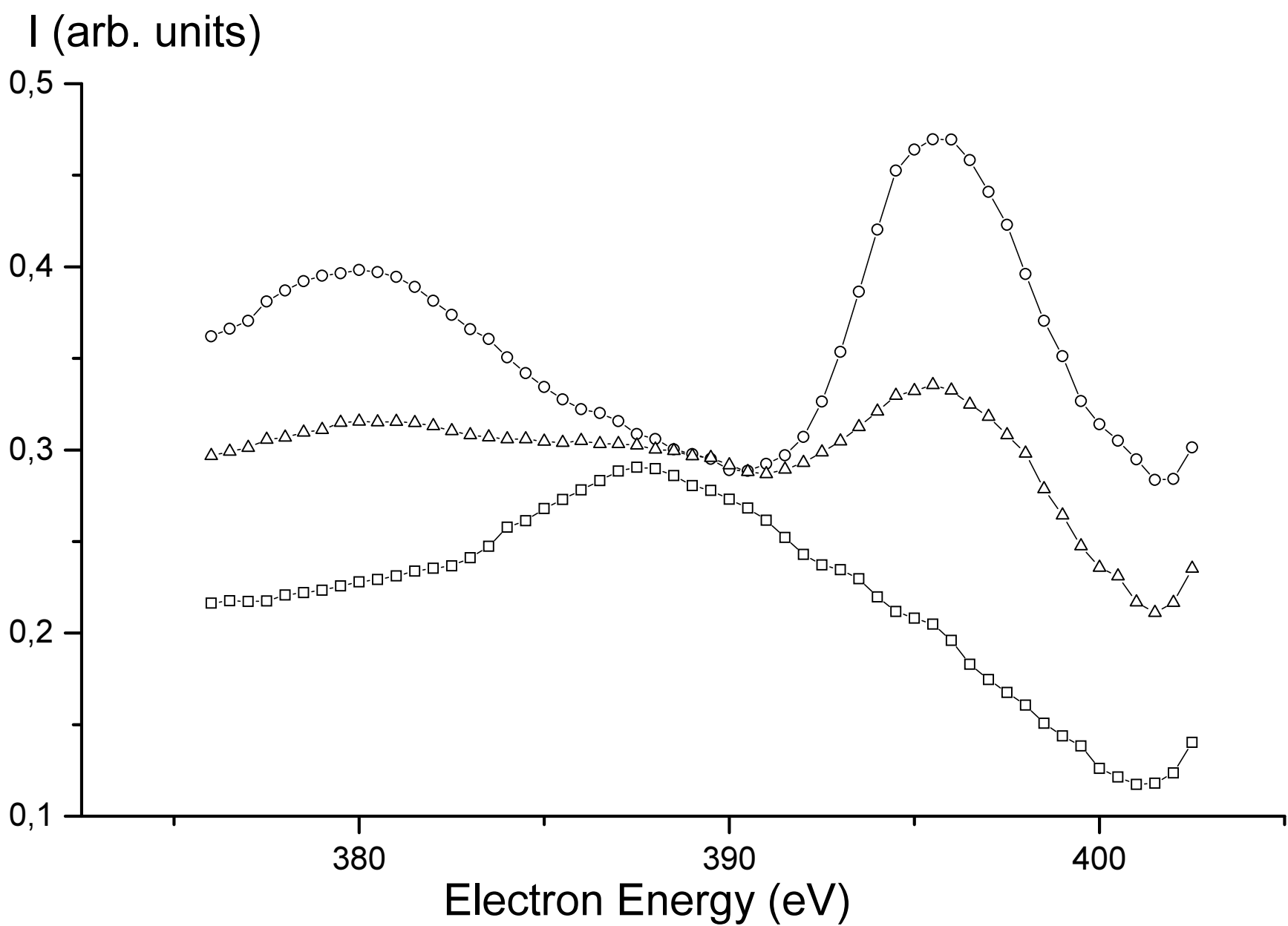
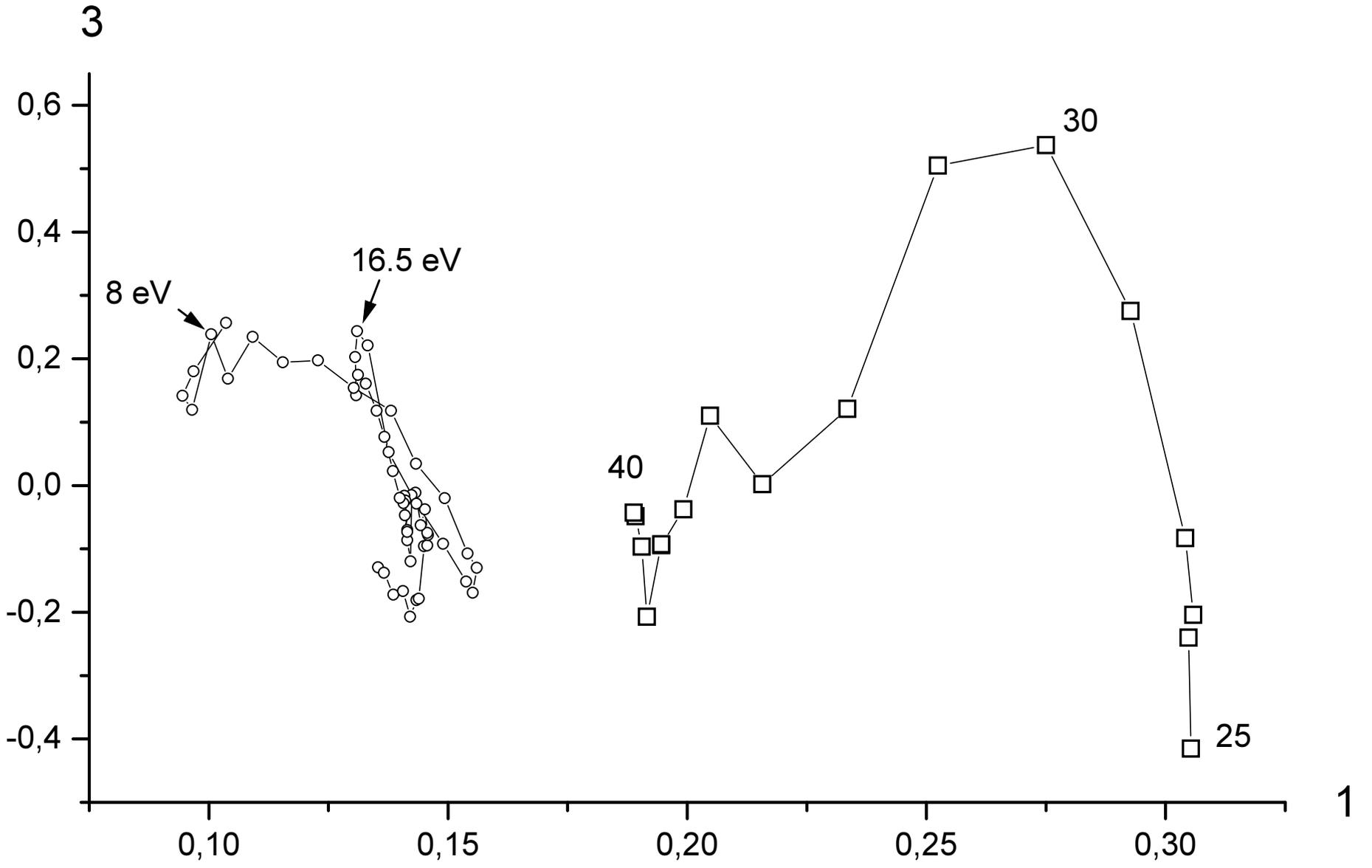
\end{document}